\documentclass[10pt, notitlepage, twoside, a4paper]{article}
\usepackage{a4wide, amssymb, epsfig, graphicx, fancyhdr}

\newcommand{\Section}[1]{\section{#1}\vspace{-8pt}}

\title{Search for $\eta$-mesic ${^4\mbox{He}}$ with WASA-at-COSY}
\author{W. Krzemie\'n \footnote{wojciech.krzemien@if.uj.edu.pl, Institute of Physics, Jagiellonian University, Cracow, Poland.} \\ for the WASA-at-COSY Collaboration}
 
\date{10.10.2010}

\begin{document}
\rhead{}
\lhead{}
\pagestyle{fancy}
\renewcommand{\headrulewidth}{0.2pt}
\renewcommand{\footrulewidth}{0.2pt}
\setlength{\parindent}{0pt}
\setlength{\parskip}{10pt}
\maketitle

\Section{Introduction}
Recently, the progress in the spectroscopy of pionic and kaonic atoms, as well as pionic and kaonic nuclei has permitted to obtain deeper insights on the meson-nucleus interaction and the in-medium behaviour of spontaneous chiral symmetry breaking~\cite{hiren}. 

Analogically to the other exotic nuclear systems, the investigation of the $\eta$-mesic nuclei would provide many interesting informations about the $\eta$-N interaction, N* in-medium properties~\cite{jido}  and would deepen our knowledge of the fundamental structure of $\eta$ meson~\cite{basssymposium}.
The $\eta$ meson is electrically neutral, therefore such a system can be formed only via the strong interaction which distinguishes it qualitatively from pionic atoms where the binding is the effect of the sum of the attractive electromagnetic force and the repulsive strong interaction. 

The search of the $\eta$~-~mesic nucleus was performed in many experiments in the past~\cite{lampf,lpi,jinr,gsi,gem,mami} and is being continued at COSY~\cite{moskalsymposium}, JINR~\cite{jinr}, J-PARC~\cite{fujiokasymposium} and MAMI~\cite{mami}. Many promising indications where reported, however, so far there is no direct experimental confirmation of the existence of mesic nucleus. In the region of the light nuclei systems such as $\eta$-$\mbox{He}$ or $\eta$-$\mbox{T}$, the observation of the strong enhancement in the total cross-section and the phase variation in the close-to-threshold region provided strong evidence to the hypothesis of the existence of a pole in the scattering matrix which can correspond to the bound state~\cite{wilkin}. 
However, as it was stated by Liu~\cite{liu3}, the theoretical predictions of width and binding energy of the $\eta$-mesic nuclei is strongly dependent on the not well known  subtreshold $\eta$-nucleon interaction. Therefore, the direct measurements which could confirm the existence of the bound state, are mandatory.

\Section{Experiment}

In June 2008 we performed a search for the $^4\mbox{He}-\eta$ bound state by measuring the excitation fun\-ction of the $dd \rightarrow ^3\mbox{He} p\pi^-$  reaction near the eta production threshold using the WASA-at-COSY detector. During the experimental run the momentum of the deuteron beam was varied continuously within each acceleration cycle
from  2.185~GeV/c to 2.400~GeV/c, crossing the kinematic threshold for the $\eta$ production in the $dd \rightarrow {^4\mbox{He}}\,\eta$ reaction at 2.336~GeV/c. 
This range of beam momenta corresponds to the variation of $^4\mbox{He}-\eta$  excess energy  from -51.4~MeV to 22~MeV .
The experimental method is based on measuring the excitation function for chosen decay channels of the ${^4\mbox{He}}-\eta$ system and a search for a resonance-like structure below the ${^4\mbox{He}}-\eta$ threshold.  The relative angle between the outgoing $p - \pi^-$ pair which  originates from the decay of the N*(1535) resonance created via  absorption of the $\eta$ meson on a nucleon in the $^4\mbox{He}$ nucleus, is 180$^{\circ}$ in the N* reference frame and is smeared by about 30$^{\circ}$ in the center-of-mass frame due to the Fermi motion of the nucleons inside the $^4\mbox{He}$ nucleus. The center-of-mass kinetic energies of the $p$ and $\pi^- $ originate from the mass difference $m_{\eta}-m_{\pi}$ and are around 50 MeV and 350 MeV, respectively.

\begin{figure}[h]
    \begin{center}
        {\includegraphics[scale=0.290]{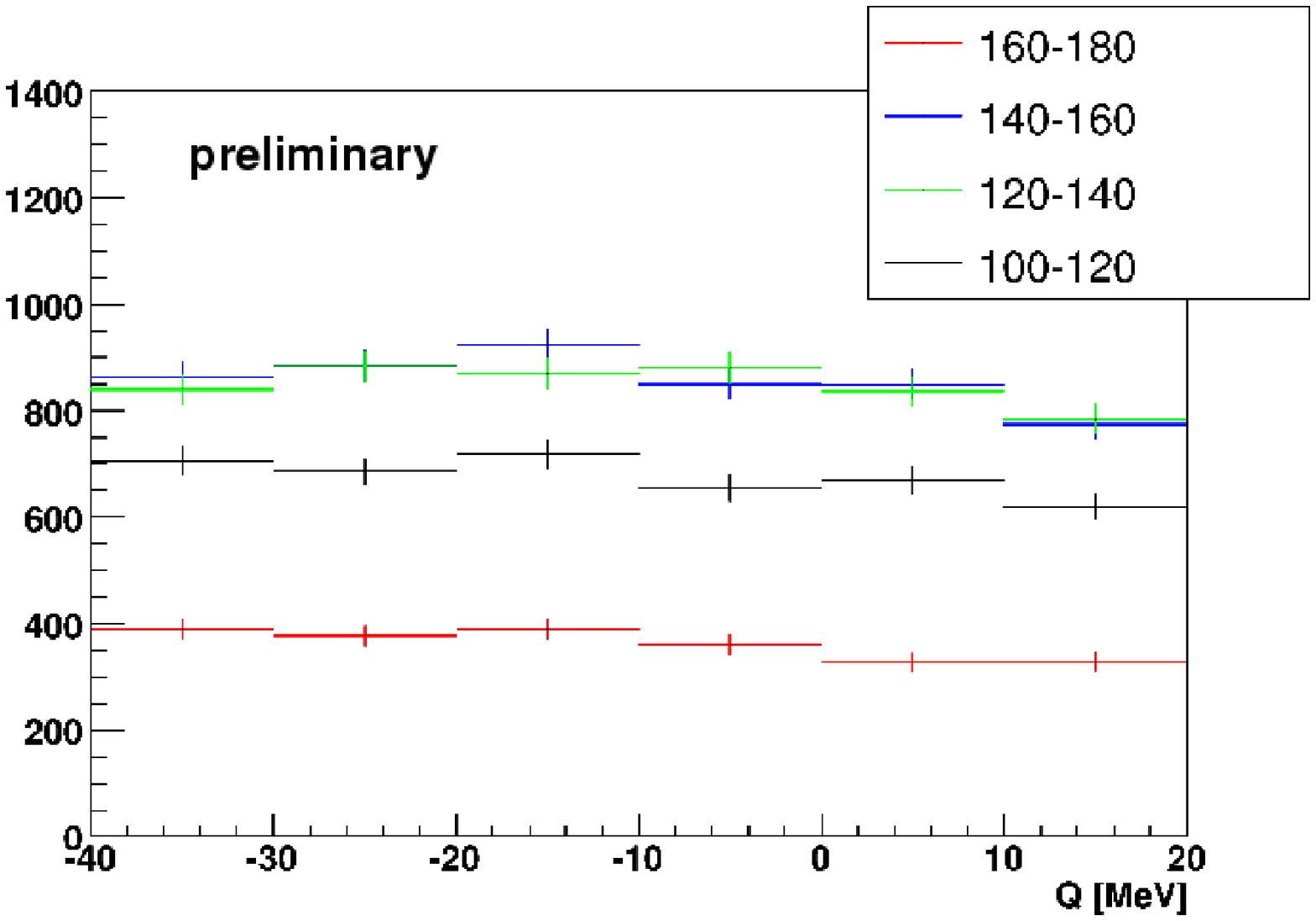}} \hfill
        {\includegraphics[scale=0.300]{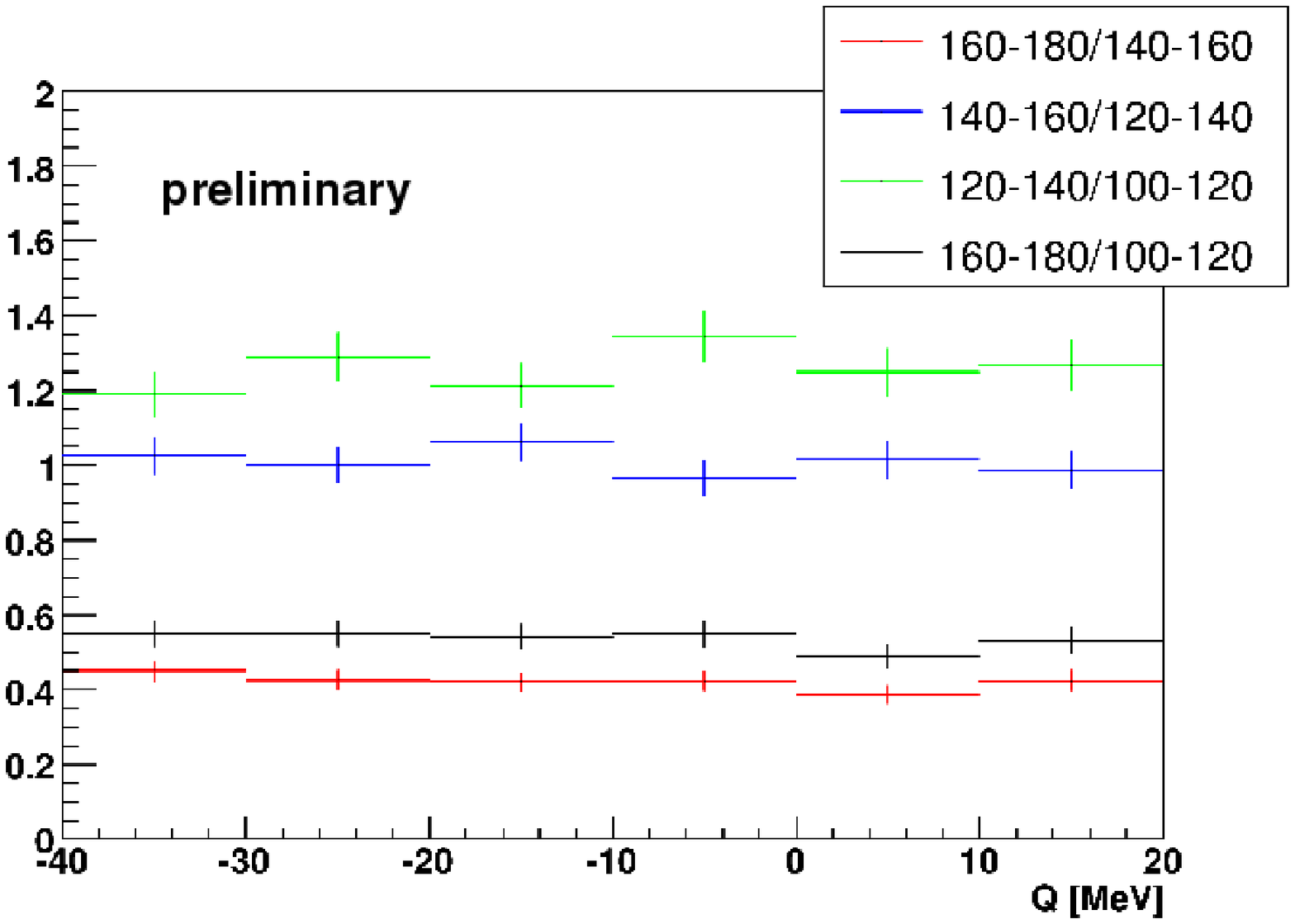}} 
        \caption{(left) Excitation function for the  $dd\to{^3\mbox{He}}p\pi$ reaction measured in the 20 degrees intervals of the 
 $\Theta_{cm}(p-\pi)$ angle. (Right) Ratio of excitation functions angular ranges as indicated in the figure. 
\label{openangle_cut_div_p}} 
    \end{center} 
\end{figure}

The Figure \ref{openangle_cut_div_p} presents the preliminary excitation function in 20 degrees intervals in $\Theta_{cm}(p-\pi)$ angle. 

The figure indicates no structure in the angular range close to the 180 degree where the signal is expected. The ratio of excitation functions from various angular ranges
is also constant as indicated in the right panel of Fig.~\ref{openangle_cut_div_p}. Therefore, taking into account the luminosity and the detector acceptance we preliminary estimated that an upper limit for the $\eta$-mesic  production via the $dd \to (\eta\,{^4\mbox{He}})_{b.s.} \to {^3\mbox{He}}\, p\, \pi^-$ reaction is equal to  about
20~nb on a one $\sigma$ level.

\Section{Outlook}

The research program of $\eta$-mesic $\mbox{He}$ with WASA-at-COSY will be continued. A two-week measurement with WASA-at-COSY for the $dd \rightarrow {^3\mbox{He}} p \pi^-$ channel is scheduled for November 2010. After two weeks of measurement with a luminosity of $4 \cdot 10^{30}$ cm$^{-2}$ s$^{-1}$, we expect a statistical sensitivity of a few nb ($\sigma$). A non-observation of this signal would significantly lower the upper limit for the existence of the bound state.

\end{document}